\documentclass[aps,prl,twocolumn,amsmath,amssymb]{revtex4-2}
\usepackage{graphicx}
\usepackage{xcolor} 
\usepackage{dcolumn}
\usepackage{bm}
\begin{document}

\preprint{APS/123-QED}

 \title{Resonantly Driven Electron Bernstein Waves in Magnetized Low-Pressure Capacitive Discharges
}
\author{Deepak Gautam$^1$}
\author{Sarveshwar Sharma$^{2,3,}$}
\email{sarvesh@ipr.res.in} \email{sarvsarvesh@gmail.com}
\author{Igor Kaganovich$^4$}
\author{Bhooshan Paradkar$^1$}

\affiliation{%
 $^1$ School of Physical Sciences, UM-DAE Centre for Excellence in Basic Sciences, University of Mumbai, Mumbai 400098, India}
\affiliation{$^2$Institute for Plasma Research, Bhat, Gandhinagar, Gujarat 382428, India}
\affiliation{$^3$Homi Bhabha National Institute, Training School Complex, Anushaktinagar, Mumbai 400094, India}
\affiliation{$^4$Princeton Plasma Physics Laboratory, Princeton, New Jersey 08543, USA 
}%




\date{\today}
\begin{abstract}
The physics of capacitively coupled plasma (CCP) discharges is investigated in a mildly magnetized regime, defined by $\mathrm{1 \leq f_{ce}/f_{rf} < 2}$, where $\mathrm{f_{ce}}$ and $\mathrm{f_{rf}}$ denote the electron cyclotron frequency and the applied radio-frequency (RF), respectively. A distinctive feature of this regime is the excitation of electron Bernstein waves (EBWs) that propagate into the bulk plasma. As the applied magnetic field increases, notable changes in the discharge characteristics occur, with EBWs observed to propagate along the plasma density gradient inside the bulk. The underlying physics of CCP operation in this regime is analyzed in detail using particle-in-cell Monte Carlo collisions (PIC-MCC) simulations.
\end{abstract}
 
\maketitle
Wave phenomena are fundamental to plasmas, governing critical processes such as heating, particle acceleration, energy transport, and instability evolution \cite{liberman,chabert2011physics,chen1984introduction}. In magnetized plasmas, a broad spectrum of electromagnetic and electrostatic wave modes can be excited depending on plasma conditions and external magnetic field configurations \cite{liberman,chabert2011physics}. These waves efficiently couple external energy into plasma and are vital in space and laboratory systems, especially where precise control of energy deposition and particle dynamics is required.\\
One such engineered system is the low-pressure RF CCP discharge, which serves as the foundation for a wide range of advanced technologies with substantial industrial and social impact \cite{liberman,chabert2011physics,chen1984introduction,makabe2006plasma}. CCPs are widely utilized in semiconductor manufacturing, surface modification, thin-film deposition, biomedical engineering, and environmental remediation \cite{coburn1979ion,gottscho1992microscopic,hopwood1992review,kong2009plasma}. In these contexts, RF plasmas facilitate processes such as ion-assisted etching and ion implantation with nanoscale precision \cite{coburn1979ion,gottscho1992microscopic,hopwood1992review,kong2009plasma,madou2011manufacturing}.\\
In recent years, significant research attention has shifted towards operating CCP discharges at very-high frequencies (VHF), typically in the 30-300 MHz range. Compared to conventional RF frequencies (e.g. 13.56 MHz), VHF driven CCP discharges exhibit distinct physical phenomena, such as strong electric field transients in the bulk plasma, dynamic sheath modulation, energetic electron beam formation, electrostatic wave excitation, field reversal, and the generation of higher harmonic components in both voltage and current waveforms \cite{sharma2016effect,upadhyay2013effect,sharma2018influence,sharma2019electric,miller2006spatial,sharma2018spatial,wilczek2018disparity,sharma2019influence,sharma2020high,simha2023kinetic,sharma2024harmonic,gozadinos2001collisionless,vender1992electron,sharma2013critical,sharma2013simulation,sharma2013simulation2,sharma2013investigation,sharma2014observation}. These phenomena strongly influence electron heating mechanisms, plasma uniformity, and process efficiency, thereby making VHF operation highly attractive for next-generation plasma-based applications.\\
Applying a static magnetic field to VHF CCP systems introduces an additional degree of freedom, enabling the excitation of wave modes typical of magnetized plasmas \cite{muller1989magnetically,lieberman1991model,hutchinson1995effects,park1997reactor,kushner2003modeling,vasenkov2004modeling,you2011role,yang2017magnetical,yang2018magnetical,sharma2018plasma,sydorenko2006particle,barnat2008rf,fan2013study,patil2022electron,zhang2021resonant,sharma2022plasma,hutchinson2002effects}. Among these, a particularly intriguing candidate is the electrostatic mode that plays a significant role in energy transport and electron heating. Unlike conventional electromagnetic waves such as the ordinary (O-mode) and extraordinary (X-mode), which are limited by cut-off conditions in overdense plasmas, EBWs are free from such constraints. This allows them to propagate and deposit energy in regions where electromagnetic waves become evanescent \cite{preinhaelter1973penetration,sugai1981mode}.\\
EBWs are excited at harmonics of the electron cyclotron frequency ($\mathrm{\omega_{ce} = eB/m_{e}}$) and are primarily driven by the perpendicular thermal motion of electrons relative to the magnetic field. These waves have been extensively studied in plasma environments such as Earth’s magnetosphere \cite{walker1993global}, solar radio bursts \cite{wu2000new}, and tokamaks, where they are used for localized heating and current drive, especially under low-collisionality conditions \cite{de1979bernstein,mcgregor2007flux,petrov1994current,urban2011survey,laqua2007electron,ram2002emission,mueck2007demonstration}.\\
Despite their proven utility in fusion research, EBWs remain largely unexplored in the context of magnetized CCP discharges, particularly those operating under low-pressure ($<$10 mTorr) and VHF conditions. In such regimes, collisional (ohmic) heating becomes increasingly ineffective due to diminished electron-neutral collision rates. This opens a promising opportunity for wave-particle interactions to significantly enhance electron heating and modify the overall plasma behavior through excitation of EBWs.\\
In this work, we present comprehensive kinetic simulations using the 1D3V PIC-MCC code, Electrostatic Direct Implicit Particle-In-Cell (EDIPIC), to study EBW excitation and propagation in magnetized, low-pressure CCPs.
The simulations reveal distinct electrostatic wave structures in the plasma bulk, which match the characteristics of Bernstein modes. These waves, propagating along the density gradient, are affected by magnetic field strength and phase-dependent electron dynamics within the bulk. By analyzing the spatiotemporal evolution of the electric field and the 2D FFT spectrum of electric field data, we identify the critical conditions under which EBWs emerge and influence bulk plasma properties. The mechanism of EBW excitation observed in magnetized CCP discharges is fundamentally different from the conventional O–X–B mode conversion process employed in conventional fusion devices. In fusion plasmas, EBWs are typically generated through externally launched electromagnetic waves that undergo successive mode conversion in strongly magnetized, inhomogeneous plasmas. In contrast, the EBWs reported here emerge intrinsically from the nonlinear dynamics of the CCP discharge itself. Specifically, the oscillating RF sheath and its nonlinear interaction with plasma electrons generate harmonic electric field components that directly excite electrostatic Bernstein modes inside the overdense plasma bulk, without the need for externally injected microwave radiation or mode-conversion schemes. This demonstrates a fundamentally new pathway for EBW generation in weakly magnetized low-temperature plasmas and highlights a previously unexplored mechanism of wave-driven power deposition in RF capacitive discharges. These findings offer a deeper physical understanding of magnetically enhanced CCPs and present a potential pathway for tailoring plasma behavior via wave–particle interaction mechanisms, which is highly relevant for future plasma-based processing technologies. A companion submission to Physical Review E \cite{LongPRE2025} provides a detailed discussion of physics in this regime.

\begin{figure*}
    \centering
    \includegraphics[width=1.0\textwidth]{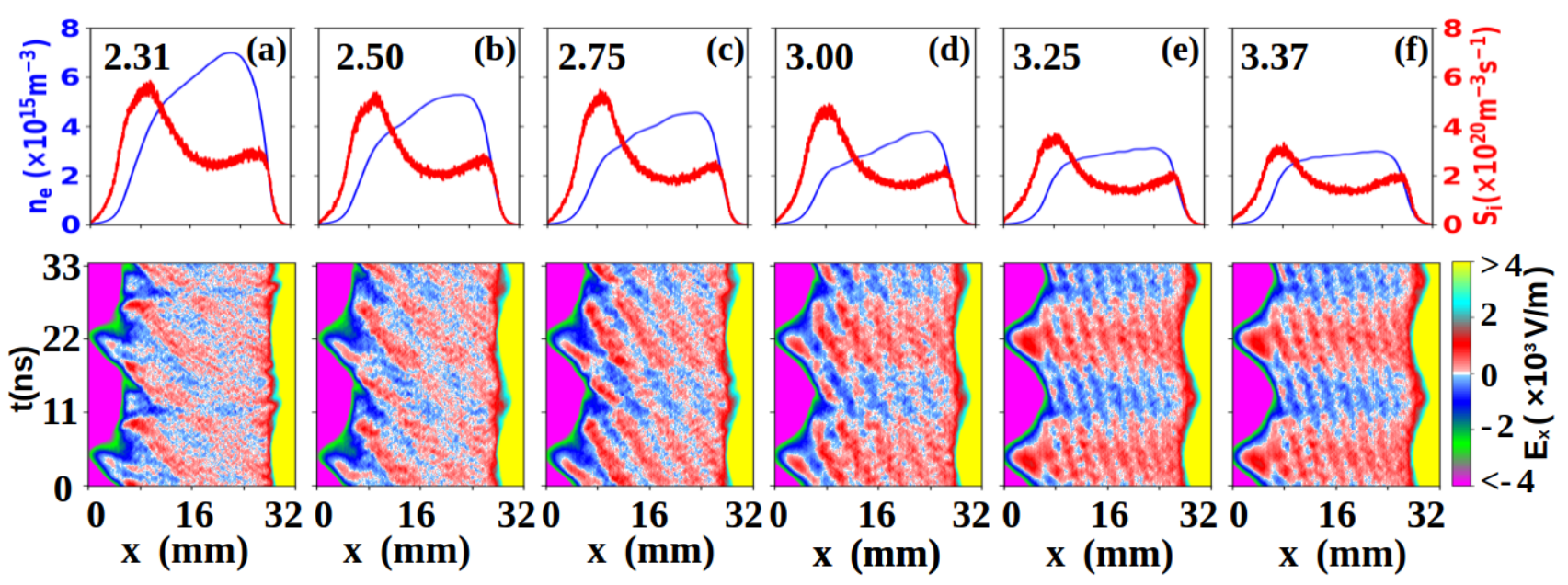} 
    \caption{The first row [(a)–(f)] presents the time-averaged electron density (blue) and ionization rate (red) for different ``r". The second row shows the spatio-temporal evolution of the electric field over the last two RF cycles at steady state for the corresponding cases. The inclined transient structures indicate a wave propagating from GE to PE.
}
    \label{combine_fig}
\end{figure*}

The magnetization of the CCP discharge is characterized by the non-dimensional parameter $r$, which is defined as 
\begin{equation}
\label{ratio}
    \mathrm{r} = \dfrac{\mathrm{2  \omega_{ce}}}{\mathrm{\omega_{rf}}}.
\end{equation}
Here, $\mathrm{\omega_{ce}}$ and $\mathrm{\omega_{rf}}$ represent the electron cyclotron frequency and the applied RF frequency, respectively. For a fixed $\omega_{rf}$, the discharge undergoes characteristic changes with increasing magnetization (or $r$) \cite{patil2022electron}. Compared to unmagnetized discharge ($r = 0$), a significant enhancement in discharge density is observed for $r \simeq 1$ due to the electron bounce cyclotron  resonance (EBCR) \cite{patil2020enhanced,patil2022electron,zhang2021resonant,wang2021magnetic,sharma2022investigating}. This is followed by a secondary peak in the discharge density near $r \simeq 2.5$. Although this secondary peak is relatively lower than the one found near $r = 1$, it suggests a qualitatively new regime of magnetized CCP discharge. Eventually, a monotonic increase in the density is found for $r > 3.5$ due to the strong magnetization of the plasma \cite{muller1989magnetically,hutchinson2002effects,fan2013study}. A schematic figure and a brief description can be found in the companion paper submitted in Physical Review E. Interestingly, there exists a possibility of exciting EBW in the CCP discharge in the range $2 < r < 4$, when $\omega_{ce}$ transitions from the fundamental to the second harmonic of $\omega_{rf}$. During this transition, the EBWs can be resonantly excited when the frequency of an oscillating electric field near the sheath-edge is comparable to the cyclotron frequency. With these considerations, we perform series of simulations with varying applied magnetic field using EDIPIC
\cite{sun2023direct,simha2023kinetic,sharma2018spatial,sydorenko2006particle,campanell2012instability,sheehan2013kinetic,campanell2013influence,carlsson2016validation,charoy20192d}. All simulations, employing a planar discharge configuration, are performed with an explicit integration scheme. The electron neutral collisions are simulated by incorporating elastic scattering, excitation, and ionization events. For ion-neutral interactions, both elastic and charge exchange collisions are considered. The collisional cross-sections for these processing are obtained from standard databases \cite{rauf1997argon,lauro2004analysis}.
 Simulations neglect metastable reactions and secondary electron emission, as their influence on discharge dynamics is weak at low pressures \cite{liberman}.

In the present study, argon at 5 mTorr is used as the background gas, with the discharge driven by a 60 MHz RF voltage and a 100 V peak. This RF voltage is applied across two parallel plate electrodes, a grounded electrode (GE) and a powered electrode (PE), which are separated by a distance of 32 mm. The PE is subjected to a time-varying voltage waveform, while the GE is maintained at zero potential. The applied voltage is mathematically expressed as
    \begin{equation}
        \label{eq1}
        \mathrm{V_{rf}(t)= V_0 \ \text{Sin}(2\pi f_{rf} t + \phi)},
    \end{equation}
where $\mathrm{V_0}$ is the peak voltage, $\mathrm{f_{rf}}$ is the RF frequency, t is time. All simulations use the initial phase angle, $\phi$ is set to zero, and effects of the external matching circuit are excluded from the model.
Since we are interested in excitation of EBWs, an external magnetic field is varied in the range $2.0 \leq r \leq 3.5$, corresponding to a field strength of 20 to 75 G. The magnetic field is applied  parallel to the electrode plates.  

All simulations are performed with grid resolution of $\Delta \mathrm{x} = \lambda_D/ 16$, where the electron Debye length ($\lambda_D$) is calculated at the initial temperature of 2 eV and the number density of $\mathrm{5\times 10^{15} \ \mathrm{m}^{-3}}$. The corresponding time step is calculated as $\Delta \mathrm{t} = \Delta \mathrm{x} / \mathrm{v_{max}}$, where $\mathrm{v_{max}}$ is set to the four-times electron thermal velocity. This timestep ensures that both $\omega_{pe}$ and $\omega_{ce}$ are well resolved.
The simulations are initialized with 400 super particles per cell, using perfectly absorbing boundary condition on electrodes. The simulations are run until steady-state conditions are reached, which typically occurs over 6000 RF cycles. 

In the regime of our interest, i.e. $2 \leq r \leq 3.5$, the discharge characteristic undergoes qualitatively significant changes with increasing magnetic field (or `$r$'). This behavior can be seen from Fig.~\ref{combine_fig}. In this regime, the symmetric discharge, when $r = 2.0$, becomes asymmetric with increasing magnetic fields \cite{yao2024spontaneous}. The asymmetry is maximum for the magnetic field corresponding to $r = 2.31$ (see Fig.~\ref{combine_fig}(a)) and slowly decreases with further increase in the magnetic field. Eventually, the discharge becomes nearly symmetric for $r \simeq 3.37$ (Fig.~\ref{combine_fig}(f)). This transition from asymmetric to symmetric discharge is shown in the top panel of Fig.~\ref{combine_fig}, where the time-averaged electron density (shown in blue) and the ionization rate (in red) are plotted for various values of `$r$'. Here, we see that the electron density peaks near the GE (at $\mathrm{x} = 32 \, \mathrm{mm}$), whereas the ionization rate maximizes near the PE (at $\mathrm{x} = 0 \, \mathrm{mm}$). This seemingly counter-intuitive behavior of the discharge is due to the enhanced electron flux from the PE side, resulting in increased ionization. In order to maintain the quasi-neutrality of the bulk plasma, the ion flux increases on the GE side, leading to peak in the time-averaged density.  This trend is demonstrated in Figure 3 of the companion paper \cite{LongPRE2025}.\\
The bottom panel of Fig.~\ref{combine_fig} shows the spatio-temporal electric field, plotted over two RF-cycles after the steady-state is reached. The panel clearly demonstrates the penetration of the RF electric field \cite{kaganovich2006revisiting} and the wave-like structures that emit from the sheath-edge of the GE and propagate down the density gradients towards the PE. The tilt of these structures in the $x-t$ plots can be used to infer their speed inside the bulk plasma. As `$r$' changes from 2.31 (Fig.~\ref{combine_fig}(a)) to 3.37 (Fig.~\ref{combine_fig}(f)), the tilt becomes steeper, indicating reduction in the propagation speed with increasing magnetic field. Apart from the tilt, we also observe that the wave-numbers ($k$) associated with these structures also increase with the increasing magnetic field (or `$r$'). For example, we see 4 distinct tilted structures for $r = 2.31$, whereas about 6 such structures are observed for $r = 2.75$ (Fig.~\ref{combine_fig}(c)). As $k$ increases with increasing `$r$', the tilted structures become closer to eventually merge into a large-scale structure, as can be seen for $r = 3.25 \, \mathrm{and} \, 3.37$ (Fig.~\ref{combine_fig}(e,f)). As the density profile becomes more and more symmetric ($r \geq 3.0$), we see gradual breaking of these structures during the half-cycles when sheath near PE is expanding and GE is collapsing. \\
\begin{figure}
    \centering
    \includegraphics[width=0.5\textwidth]{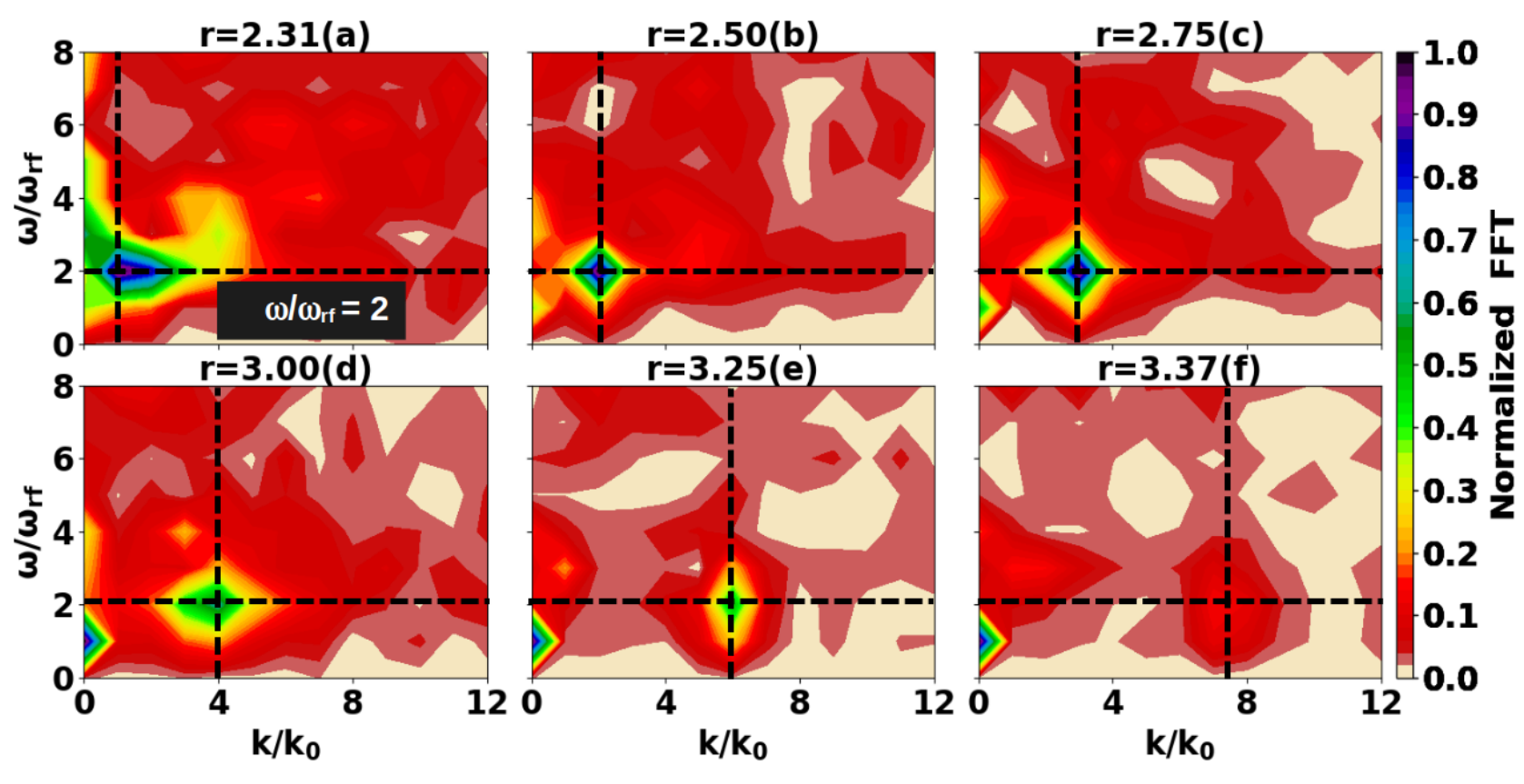} 
    \caption{Normalized 2D FFT spectrum of the electric field, showing frequency distribution and magnitudes. Dotted lines indicate the second harmonic ($\mathrm{\omega/\omega_{rf}=2}$) and the corresponding normalized wavenumber ($\mathrm{k/k_0}$), with $\mathrm{k_0}$ defined by inverse of bulk plasma length.}
    \label{2d_fft}
\end{figure}

\begin{figure}
    \centering
    \includegraphics[width=0.5\textwidth]{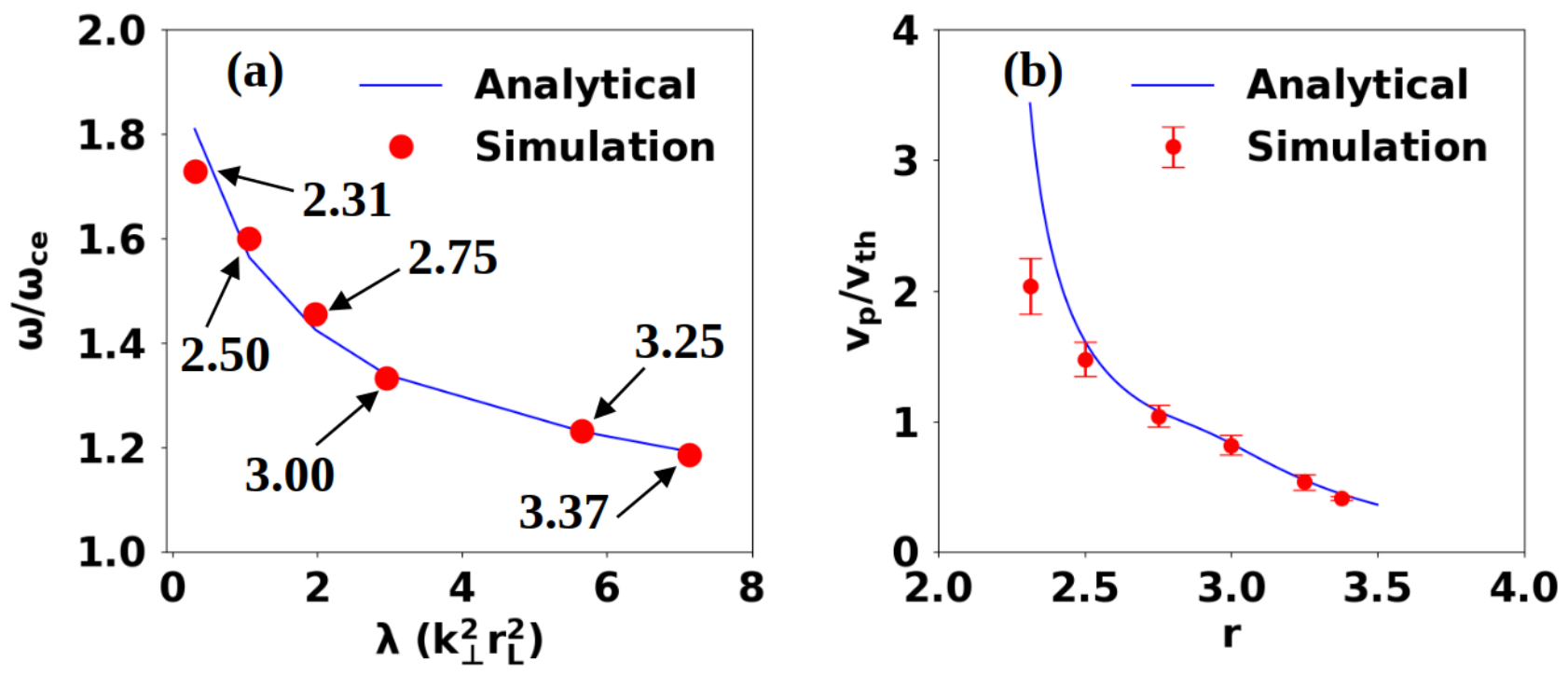} 
    \caption{(a) EBW dispersion relation, showing the shift of the dominant frequency from the second harmonic ($2\,\omega_{\mathrm{ce}}$) to the fundamental harmonic of the cyclotron frequency as $\mathrm{k_{\perp}^{2} r_{L}^{2}}$ increases. (b) Comparison of normalized phase velocity ($v_{p}/v_{\mathrm{th}}$), obtained from the tilt of wave structures in Fig.~\ref{combine_fig} with analytical expression derived from Eq.~\ref{neg_sol} for various values of `r'. The electron temperature is fixed at 2 eV. 
}
    \label{disp_vp}
\end{figure}

We construct the dispersion relation associated with the waves, seen in Fig.~\ref{combine_fig}, by performing a 2D-FFT analysis on the spatio-temporal electric field. Here, we only consider the fields within the bulk plasma region between $\mathrm{8 \ mm \ to \ 24 \ mm}$. The results are shown in Fig.~\ref{2d_fft} where the frequency ($\omega$) and the wavenumber ($k$), obtained from spatio-temporal data, are shown for each value of `$r$' shown in Fig.~\ref{combine_fig}. The frequencies are normalized with the applied RF frequency, i.e. $\mathrm{f_{rf}=60 \ MHz}$, and the wave numbers $\mathrm{k}$ are normalized with the inverse of the bulk plasma length, $\mathrm{k_{0}=1/0.016 \approx 62.5 \ m^{-1}}$. In this figure, we see a distinct peak in the second harmonic of the RF frequency ($\omega/\omega_{rf} = 2$) at 120 MHz. With increasing magnetic field strength, the peak shifts towards higher $k$ values, consistent with the trend shown in Fig.~\ref{combine_fig}. Note that for $r = 2.31, \, 2.50, \, 2.75$ (Fig.~\ref{2d_fft}(a-c)), the second harmonic of the RF frequency is dominant, but at higher magnetic fields, corresponding to $r = 3.00, \, 3.25, \, 3.37$ (Fig.~\ref{2d_fft}(d-f)), the fundamental mode (60 MHz) dominates, as can be seen from figure 7 of the companion Physical Review E paper \cite{LongPRE2025}.  The dispersion relation is now constructed numerically from the simulation data using the values $\omega$ and $k$ at the intersections of the horizontal and vertical lines shown in the figure.
To confirm that the waves reported in Fig.~\ref{combine_fig} are indeed EBWs, we compare the numerically obtained dispersion relation from Fig.~\ref{2d_fft} with the analytical expression. The results of this comparison are shown in Fig.~\ref{disp_vp}. \\
Theoretically, the relation between frequency ($\omega$) and wavenumber ($\mathrm{k_{\perp}}$) for an EBW \cite{bellan2008fundamentals} is obtained by  
\begin{equation}
\label{e_disp}
\mathrm{D\left(k_{\perp},\omega\right)= 1 - \dfrac{2e^{-\lambda}}{\lambda} \sum_{n=1}^{\infty} \dfrac{n^{2}\omega_{pe}^{2}}{\omega^{2}-n^{2}\omega_{ce}^{2}} \text{I}_{\text{n}}(\lambda) = 0}.
\end{equation}
Here, $\text{I}_{\text{n}}$ is the modified Bessel function of order $\text{n}$ and $\lambda$ is the Larmor parameter defined as
\begin{equation}
\label{lambda}
\mathrm{\lambda =k_{\perp}^{2}  r_{L}^{2}}.
\end{equation}
Note that $\lambda$ represents the square of the wavelength of EBW, normalized by the electron Larmor radius $\mathrm{r_{L}}$. For the fundamental harmonic of EBW, we consider only the first two terms in the series expansion of Eq.~\ref{e_disp}, which gives the following.
\begin{equation*}
\begin{split}
1 &= \dfrac{2e^{-\lambda}}{\lambda} \left[ \dfrac{\omega_{pe}^{2}}{\omega^{2}-\omega_{ce}^{2}} \text{I}_{1}(\lambda) + \dfrac{4 \omega_{pe}^{2}}{\omega^{2}-4 \omega_{ce}^{2}} \text{I}_{2}(\lambda) \right].
\end{split}
\end{equation*}
Solving quadratic in $\omega^2$, we get 

\begin{equation}
\label{neg_sol}
\frac{\omega^2}{\omega_{ce}^2} = \dfrac{5+\alpha}{2} - \dfrac{1}{2} \sqrt{9 + 10\alpha + \alpha^{2} - 16\beta}.
\end{equation}
Here,
\begin{equation*}
\begin{split}
\alpha &= \dfrac{2e^{-\lambda}}{\lambda} \dfrac{\omega_{pe}^{2}}{\omega_{ce}^{2}} \left[ \text{I}_{1}(\lambda) + 4 \ \text{I}_{2}(\lambda) \right], \\
\beta &= \dfrac{2e^{-\lambda}}{\lambda} \dfrac{\omega_{pe}^{2}}{\omega_{ce}^{2}} \left[ \text{I}_{1}(\lambda) + \text{I}_{2}(\lambda) \right]
\end{split}
\end{equation*}
In the long-wavelength limit ($\lambda \rightarrow 0$), we get $\mathrm{\alpha \simeq \beta \simeq \dfrac{\omega_{pe}^{2}}{\omega_{ce}^{2}}>>1}$ and $\mathrm{\omega \simeq 2 \omega_{ce}}$. On the other hand, in the short wavelength limit ($\mathrm{\lambda \rightarrow \infty}$), Eq.~\ref{neg_sol} reduces to $\mathrm{\omega \simeq \omega_{ce}}$. The dependence of $\mathrm{\omega/\omega_{ce}}$ on $\mathrm{\lambda = \left(k_{\perp}r_L\right)^2}$ for the fundamental harmonic ($\mathrm{n = 1}$) of EBW, obtained from Eq.~\ref{neg_sol} is plotted in Fig.~\ref{disp_vp}(a). In this figure, the blue line represents the analytical solution, whereas the solid red circles are the numerically obtained modes from the 2D FFT analysis, presented in Fig.~\ref{2d_fft}. Note that in the analytical expression, we have taken $\omega_{pe}$ corresponding to the average bulk plasma density and $T_e = 2 \, \mathrm{eV}$. The phase velocity (normalized by the thermal velocity of electrons) of the waves can also be obtained by dividing Eq.~\ref{neg_sol} by $\lambda$. This phase velocity can be compared with the propagation speed of the waves obtained from the tilt in the wave-like structures in the spatio-temporal electric field. This comparison, shown in Fig.~\ref{disp_vp}(b), demonstrates excellent agreement in the propagation speed obtained from numerical simulations and analytical expression. This unequivocally demonstrates the propagation of EBW in magnetized CCP discharges.  \\ 
Using Eq.~\ref{neg_sol}, we can also obtain the expression for the ratio of group velocity ($v_{g}$) to phase velocity ($v_{p}$) of EBW in the following form:
\begin{equation}
    \label{group_vel}
    \frac{v_{g}}{v_{p}} = \frac{2\lambda}{\left(\omega/\omega_{ce}\right)}\frac{\partial \left(\omega/\omega_{ce}\right)}{\partial \lambda}.
\end{equation}
From the dispersion relation demonstrated in Fig.~\ref{disp_vp}, we see that $\partial \left(\omega/\omega_{ce}\right)/\partial \lambda < 0$. This shows that in this regime, the EBWs are backward propagating waves i.e. the group velocity and phase velocity are in the opposite direction. Therefore, although in Fig.~\ref{combine_fig}, the phase fronts appear to move from GE to PE, EBWs are generated near the PE carrying energy towards the GE.

\begin{figure}
    \centering
     \includegraphics[width=0.5\textwidth]{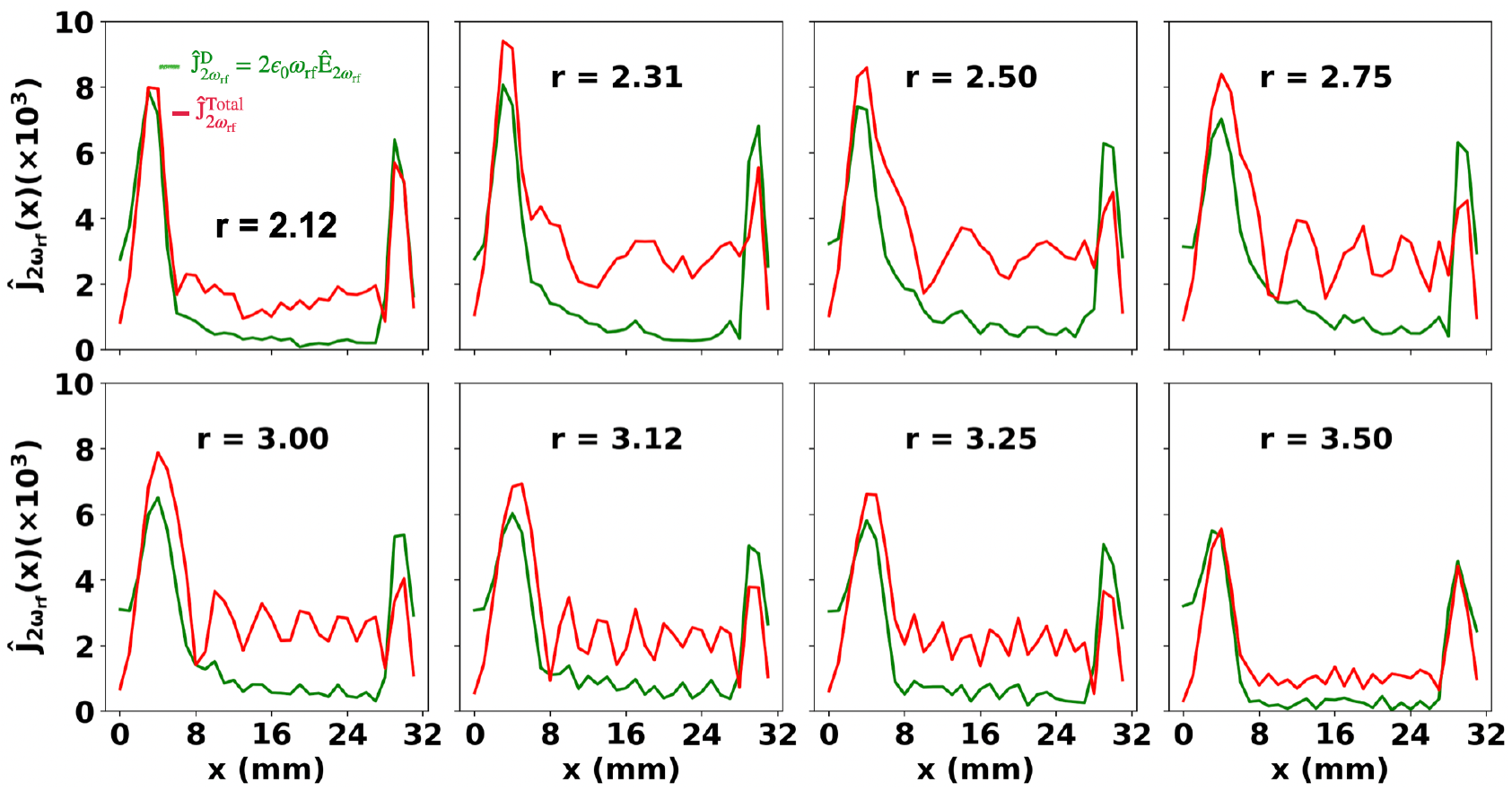}
    \caption{The spatial profile of second harmonic component of total electron current (red color). The green curve represents the second harmonic content in the displacement current estimated second harmonic of the electric field. Note that the displacement current due to sheath oscillations acts like a driver for EBW excitation as per Eq.~\ref{cary_PRE_2007_eq3}.}
    \label{je_second_harmonic}
\end{figure}

To understand the mechanism of EBW generation, we consider the excitation of EBW through the external driving current $j_e$. The spatio-temporal Fourier transform of the electric field associated with EBW ($\tilde{E}(k_{\perp},\omega)$) can be described by the equation in the form \cite{cary2007wave}
\begin{equation}
    \label{cary_PRE_2007_eq3}
    i\omega \epsilon_0 D(k_{\perp},\omega) \tilde{E}(k_{\perp},\omega) = \tilde{j_e}(k_{\perp},\omega).
\end{equation}
Here, $D(k_{\perp},\omega)$ is the plasma dielectric response described by Eq.~\ref{e_disp} and $\tilde{j_e}(k_{\perp},\omega)$ is the Fourier transform of any current that is not included in the dielectric response. In case of CCP discharges, the displacement current due to the time varying sheath electric field ($E_{sh}$) plays the role of driving current, i.e. $j_e \simeq \epsilon_0 \dfrac{\partial E_{sh}}{\partial t}$. Therefore, non-linear sheath oscillations can resonantly drive the waves as long as the driving current has appropriate harmonics to match the frequency and wave-number of the excited EBW mode. \\

From the dispersion relation shown in Fig.~\ref{disp_vp}, we know that the fundamental EBW mode, corresponding to $n = 1$ in Eq.~\ref{e_disp}, must lie close to the second harmonic of the RF frequency, i.e., $\omega \simeq 2 \omega_{rf}$. Therefore, the presence of the second harmonic in the driving current can resonantly excite the fundamental EBW mode. This can be seen in Fig.~\ref{je_second_harmonic}, where we have plotted the second harmonic of the total electron current density ($j_e$) along the discharge length in red. This quantity, denoted by $\hat{j}_{2\omega_{rf}}$, has contributions from both conduction and displacement currents. The contribution of the displacement current, estimated as $\hat{j}^D_{2\omega_{rf}}=2\omega_{rf}\epsilon_0\hat{E}_{2\omega_{rf}}$, is shown in green color. As expected, we find that the current near the sheath region is dominated by the displacement current associated with the sheath oscillations. Therefore, resonant excitation of the fundamental mode of EBW can be attributed to the presence of a second harmonic in the electron current density near the sheath region, caused by sheath oscillations. Since the electric field of the oscillating sheath influences the motion of the electrons near the sheath region, we have performed detailed analysis of the trajectories of test particles placed near the sheath region. The Fourier analysis of these trajectories also confirms that these electrons dominantly oscillate with twice the RF frequency to excite the fundamental EBW mode. Detailed discussion of test particle trajectories can be found in the companion paper submitted to Physical Review E\cite{gautam2025simulation}.      \\

To address the influence of non-uniform plasma density ($n_e(x)$) on evolution of EBWs, we observe that along the distance $s$ of the wave trajectory, $k_{\perp}$ and $x$ act like momentum and position, respectively. Therefore, EBW evolution in the presence of non-uniform plasma is described by Hamilton's equations, i.e. $dk_{\perp}/ds = -\partial D/\partial x ; \,\, \,dx/ds = \partial D/\partial k_{\perp}$ \cite{cary2007wave,xiang2008second}. This leads to a wave momentum evolution in the form
\begin{equation}
    \frac{d k_{\perp}}{ds} = \left(\dfrac{2e^{-\lambda}}{\lambda} \sum_{n=1}^{\infty} \dfrac{n^{2}\omega_{pe}^{2}(x_0)}{\omega^{2}-n^{2}\omega_{ce}^{2}} \text{I}_{\text{n}}(\lambda)\right) \left(\frac{1}{n_0}\frac{d n_e}{dx}\right), 
\end{equation}
where $n_0$ is the reference electron density at the location $x_0$. Therefore, the wave gains (losses) its momentum along the positive (negative) density gradient. This explains the preferred direction of propagation of EBW from low density to high density side. This description is consistent with estimates of the X-B conversion efficiency, described in terms of the Budden parameter \cite{budden1988propagation}, which shows that the excitation of EBWs is preferable from the low density side \cite{laqua2007electron,ram2002emission}. It should be noted that, in this regime of operation, EBWs are backward propagating waves having group and phase velocities in the opposite direction. Thus, although the waves are traveling from the powered electrode to grounded electrode in the direction of group velocity, the phases appear to advance in the opposite direction (from GE to PE). This is consistent with Fig.~\ref{combine_fig}, where distinctive wave structures appear to move from GE to PE. \\

The excitation and propagation of EBWs in low temperature devices open new avenues for heating bulk plasma in the low pressure regime where collisionless heating is ineffective. This mechanism offers an alternate pathway to conventional stochastic heating mechanism, which is typically dominantly present only in the sheath region. Detail analysis of time-average electron heating, obtained from PIC simulations, clearly demonstrates enhancement in the bulk electron heating with increase in the EBW activity. These details are given in the companion paper submitted to Physical Review E\cite{gautam2025simulation}.

In summary, in this letter, we demonstrate the resonant excitation of fundamental mode of EBWs by appropriately tuning the strength of the magnetic field. These waves are found to be excited near the PE in the operating range $2.3  < r < 3.5$ where the second harmonic of RF is comparable to the frequency of the fundamental mode of EBW. The presence of a density gradient helps propagate these waves within the bulk plasma \cite{sugaya1971parametric,ram2000excitation}. Excitation of such waves in CCPs offers new plasma heating pathways when collisional and stochastic mechanisms are inefficient, especially at low pressures.\\

Acknowledgments: IDK’s research was supported by the US DOE, Office of Fusion Energy Science under the DE-AC02-09CH11466 contract, as part of the Princeton Collaborative Low Temperature Plasma Research Facility (PCRF). The Authors also acknowledge the NEUMANN (CEBS) and ANTYA (IPR) HPC facilities.

\bibliographystyle{apsrev4-2}
\bibliography{apssamp}

\end{document}